\documentclass{article}
\usepackage{pdfx}
\usepackage{microtype}
\usepackage{graphicx}
\usepackage{subfigure}
\usepackage{booktabs} % for professional tables

\usepackage{hyperref}

\newcommand{\eat}[1]{}

\usepackage[accepted]{icml2023}

\usepackage{amsmath}
\usepackage{amssymb}
\usepackage{mathtools}
\usepackage{amsthm}

\usepackage[capitalize,noabbrev]{cleveref}

\usepackage{empheq}

\theoremstyle{plain}

\theoremstyle{definition}

\theoremstyle{remark}

\usepackage[textsize=tiny]{todonotes}

\icmltitlerunning{IncDSI: Incrementally Updatable Document Retrieval}

\begin{document}

\twocolumn[
\icmltitle{IncDSI: Incrementally Updatable Document Retrieval}



\icmlsetsymbol{equal}{*}

\begin{icmlauthorlist}
\icmlauthor{Varsha Kishore}{sch}
\icmlauthor{Chao Wan}{sch}
\icmlauthor{Justin Lovelace}{sch}
\icmlauthor{Yoav Artzi}{sch}
\icmlauthor{Kilian Q. Weinberger}{sch}
\end{icmlauthorlist}

\icmlaffiliation{sch}{School of Computer Science, Cornell University, Ithaca, USA}

\icmlcorrespondingauthor{Varsha Kishore}{vk352@cornell.edu}
\icmlcorrespondingauthor{Justin Lovelace}{jl3353@cornell.edu}

\icmlkeywords{Machine Learning, ICML}

\vskip 0.3in
]



\printAffiliationsAndNotice{}  % leave blank if no need to mention equal contribution

% macros
\newcommand{\method}{IncDSI}

\begin{abstract}
Differentiable Search Index is a recently proposed paradigm for document retrieval, that encodes information about a corpus of documents within the parameters of a neural network and directly maps queries to corresponding documents. These models have achieved state-of-the-art performances for document retrieval across many benchmarks. These kinds of models have a significant limitation: it is not easy to add new documents after a model is trained. We propose IncDSI, a method to add documents in real time (about 20-50ms per document), without retraining the model on the entire dataset (or even parts thereof). Instead we formulate the addition of documents as a constrained optimization problem that makes minimal changes to the network parameters. Although orders of magnitude faster, our approach is competitive with re-training the model on the whole dataset and enables the development of document retrieval systems that can be updated with new information in real-time. Our code for IncDSI is available at \href{https://github.com/varshakishore/IncDSI}{https://github.com/varshakishore/IncDSI}.
\end{abstract}

\section{Introduction}
\label{sec:intro}
Information retrieval (IR) systems map user queries, often expressed in natural language, to relevant documents. 
They are the core technology underlying search engines, and are only becoming more critical as the information available to users grows in 
complexity and volume. 
Current retrieval methods largely align with one of two paradigms.
The dual encoder methods train separate encoders for queries and documents that map the two into a shared embedding space. 
The training loss encourages that queries are closest to their respective target documents~\citep{karpukhin2020dense,xiong2020approximate} and one can  perform retrieval by conducting a nearest neighbor search given the query and document embeddings. 
The other paradigm that is gaining significant interest recently is differentiable search indexing~\citep[DSI;][]{tay2022transformer}, in which all information about a collection of documents is encoded in the parameters of a neural network model. Given a query, the model directly returns the ID of the relevant document, either via classification over all IDs or by generating the ID with a decoder. 

The two paradigms are quite different and have complementary advantages. 
It is straightforward to add new documents to dual encoder systems by mapping them into the joined space using the trained document encoder and including the resulting embedding vectors in the nearest neighbor search. DSI systems, on the other hand, shine in offering higher flexibility to learn the retrieval encoding of a document. Here, documents are not encoded through a shared encoder, but instead their implicit representation (i.e., within the network parameters) is induced during training. 
DSI methods are also relatively simple, consisting of a single unified model instead of different encoders and search procedures; DSI models perform retrieval with a single forward pass. However, DSI systems are harder to extend to new documents. Naively training the model with new document risks catastrophic forgetting of existing documents~\cite{mccloskey1989catastrophic, toneva2018empirical, mehta2022dsi++} and retraining on old and new data on a regular basis is costly.

\begin{figure*}
    \centering
    \includegraphics[width=0.9\textwidth]{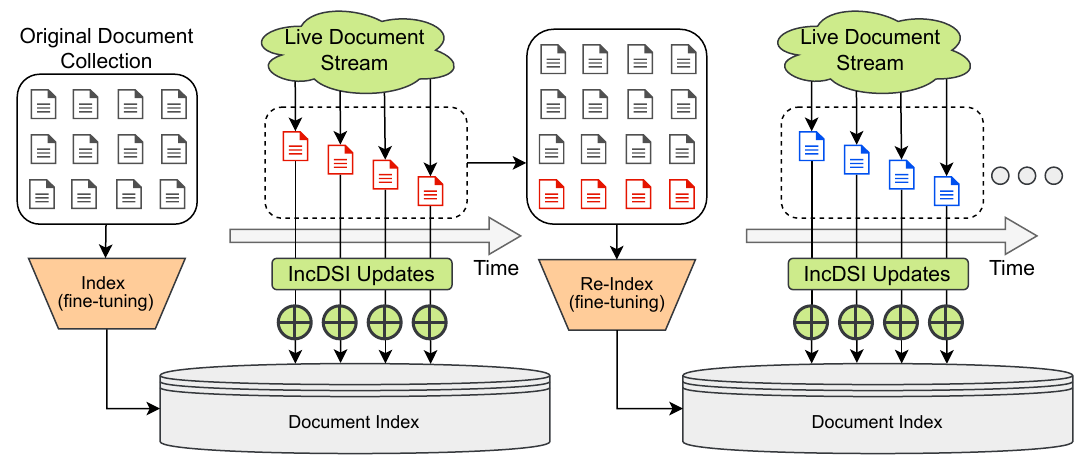}
    \caption{Overview of our proposed setting. IncDSI can index incoming documents immediately and begin serving them to users.}
    \label{fig:overview}
\end{figure*}

Most search engines retrieve documents from dynamic corpora that can grow over time. Consider a search engine for arXiv papers or social media, for instance. As new documents are uploaded, they should become available as soon as possible---ideally in real time.  \autoref{fig:overview} illustrates the setting where a document retrieval model is first trained on an initial set of documents, after which new documents arrive and must be incorporated into the document index as soon as possible. Although a DSI system can be retrained periodically, its extension to the real-time setting has so far remained an open problem. 
% In this paper, we study the problem of making DSI amenable to such settings
% building a flexible DSI-based retrieval system that can efficiently index new documents until the next time the model can be re-trained. 

We develop \emph{\method}, an approach that allows for rapidly adding new documents to a trained DSI model, while preserving the unmatched flexibility and performance of such models. 
Although in DSI the retrieval process happens inside the neural network, it is possible to formulate a constrained optimization problem that allows us to update and extend the number of document classes without 
retraining. 

Our approach leverages the fact that DSI networks have two main components: an encoder and a linear classification layer. The encoder embeds the queries and documents in a joint representation space, and the classification layer can be viewed as a matrix where each row corresponds to a \textit{document vector}. Performing classification by finding the document vector that has maximal inner product with an embedded query is effectively a nearest neighbor search of the query and the document vectors. 
This is akin to the dual encoder setup, the main difference being that the document class embeddings (i.e. the document vectors) are not the output of a document encoder, but are instead \textit{independently} learned. 
%by the insight that non-generative DSI models~\citep{tay2022transformer}, which predict a distribution over documents, conclude with a linear classification layer. Here, each document \emph{class} is represented as a learned vector. 
%This final pre-softmax layer is used to project the internal representation to a distribution over documents that is then normalized by using a softmax. All model computations that precede this part compute a query representation, which is then multiplied against the document vectors to compute per-document similarities. 
%We propose to detach the document representations to independent vectors, giving us a document space similar to dual encoder approaches. 
This independence allows us to formulate adding a new document as a \emph{constrained optimization} problem that aims to find the optimal document vector for a new document. 
The independence also guarantees that this process does not modify any other existing document vectors and does not require broader updates to the query encoder. 

% \ya{need to refer to fig1 somewhere above, where it's most useful to illustrate our scenario/approach}

We evaluate our approach by incrementally adding up to 10k documents to a trained retrieval model, evaluating both retrieval performance and the speed of adding documents. 
%, an order of magnitude faster when compared to retraining the model with the new document. 
Compared to retraining the model with the new documents,  \method{} retains retrieval performance on old documents while simultaneously achieving comparable performance on the new documents. It also has a significant advantage: \method{} is extremely fast, and only requires about 50 milliseconds to add a new document. Our code for \method{} is available at \href{https://github.com/varshakishore/IncDSI}{https://github.com/varshakishore/IncDSI}.

\section{Related work}
\label{sec:related_work}
\paragraph{Sparse and Dense retrieval methods.}
Document retrieval comprises of two main tasks- 1) \emph{Indexing}, during which document representations are learned and 2) \emph{retrieval}, during which the right document is found for a given query. Early approaches made use of sparse document and query representations due to their simplicity and effectiveness~\cite{blanco2012graph,rousseau2013graph,zheng2015learning, guo2016deep,robertson1995okapi}. However, these methods often fail to capture rich semantic connections between documents and queries. Dense retrieval methods leverage the power of neural networks to learn dense representations of documents and queries in low dimensional space. The most common dense retrieval methods use biencoders to learn to encode documents and queries such that the queries are close to their corresponding documents. During retrieval, for any given query, documents are  retrieved by using Approximate Nearest Neighbor (ANN) search~\cite{xiong2020approximate,dehghani2017neural}. \citet{karpukhin2020dense} present DPR, which is a BERT-based biencoder, trained using contrastive loss with in-batch negatives. Improving upon this work, many others explore efficient negative sampling strategies to improve the contrastive loss performance~\cite{xiong2020approximate, gao2021complement}. ANCE~\cite{xiong2020approximate} is trained with two simultaneous processes---the first refreshes the document and query embeddings periodically and the second uses the latest embeddings to find hard negatives. Cross encoders, another class of dense retrieval methods, encode queries and documents together, in order to better model the interaction between them~\cite{nogueira2019multi, qu2020rocketqa, khattab2020colbert, luan2021sparse}.

% \citet{ni2021large} shows that scaling up the language model that is used as encoder makes the model more generalizable to out-of-domain retrieval tasks.

\paragraph{End-to-end Retrieval.}
In contrast with dense dual-encoder based retrieval approaches, which perform indexing and retrieval in two separate stages, DSI~\cite{tay2022transformer} aims to combine the two stages in an end-to-end manner. For indexing, a nerual network with parameters $\theta$ is trained to map document text to corresponding document identifiers (docids). For retrieval, the neural network is trained to map user queries to docids. 
These two tasks are simultaneously learned. Document ids can either be auto-regressively generated (string ids) or produced by a dot product with a classification layer (atomic ids). Unlike parametric dense-retrieval methods, DSI is non-parametric and document specific parameters are learned. 

Many other methods build on DSI and use other techniques to further improve the model performance. \citet{wang2022neural} use generated queries and a novel auto-regressive decoder architecture to improve the DSI performance. They prepend each digit in the docid with a position number and propose a Prefix-Aware Weight Adaptive decoder. \citet{zhou2022ultron} use keyword based and semantic based docids to index the documents.

\paragraph{Query Generation in document retrieval.} Recent work has shown that using queries from a query generation model, in addition to the first few tokens of a document, to obtain its representations improves the results for document retrieval. This is because in traditional retrieval there is a mismatch between the two objectives of indexing and retrieval. \citet{zhuang2022bridging} show that performing indexing with generated queries significantly improves retrieval results. Similarly, \citet{wang2022neural} also show that using generated queries boosts the performance of neural corpus indexer (NCI), which is a sequence to sequence retrieval and is explained in the paragraph above. \citet{bonifacio2022inpars} consider settings where queries are not available for training retrieval models. They show that generated queries are not only useful for indexing but are also useful for retrieval when human queries are unavailable; in these settings they can be used in place of human queries. \citet{bonifacio2022inpars} prompt a large language model with a few document-query pairs to generate additional synthetic queries, which are then used to train information retrieval systems. 

\paragraph{Preventing forgetting.}
One of the biggest challenges in continual learning is \textit{catastrophic forgetting}~\cite{parisi2019continual}, a phenomenon in which old data is forgotten as a model is trained on new data. Alleviating forgetting is an active research area~\cite{kirkpatrick2017overcoming, riemer2018learning, lee2017overcoming}, in which memory-based approaches are popular ~\cite{hayes2019memory, isele2018selective, lopez2017gradient,chaudhry2018efficient, rolnick2019experience, aljundi2018memory}. \citet{chaudhry2019tiny} shows that repeating even a small part of old training data while the model is trained on new data can reduce forgetting to some extent. This technique is also applied in  \citet{mehta2022dsi++} where they use both generated and natural queries from old documents while training on new documents. Apart from using generated queries, they also apply Sharpness-Aware Minimization ~\cite{foret2020sharpness} in their training objective to optimize for a flatter loss basin instead of a minimal but potentially sharp loss. This method is shown to help alleviate forgetting~\cite{foret2020sharpness}.

\section{Problem Setup and Notation}
We aim to have an up-to-date real time retrieval model that can be quickly and efficiently updated with information from new documents. At any given time both queries from old and new documents must correctly be mapped to their corresponding documents. This streaming setting is pictorially shown in \autoref{fig:overview}.

Our method, \method{}, broadly has two different stages. In the first stage, a document retrieval model $M^0$ is trained on an initial set of of documents $D^0=\{d_1, \cdots , d_n\}$. Each of these documents has some number of associated queries that are used in training and we denote $\mathbf{q}_{i,{j}}$ to be the $i^{th}$ query associated with document $j$. These queries can either be user queries or queries from a query generation model; both are used in the same manner in our method.  

In the second stage, additional documents become available in a streaming fashion. As each new document becomes available, the retrieval model is updated to include it. We denote the new documents as $D' = \{d_{n\!+\!1}, d_{n\!+\!2}, \cdots \}$ and use $M^t$ to refer to the updated model after $t$ new documents have been added. Like with the initial documents, we also have some variable number of queries $\{\mathbf{q}_{1,{n\!+\!t}}, \mathbf{q}_{2,{n\!+\!t}}, \cdots\}$ corresponding to each new document $d_{n\!+\!t}$.

\begin{figure*}
    \centering
    \includegraphics[width=\textwidth]{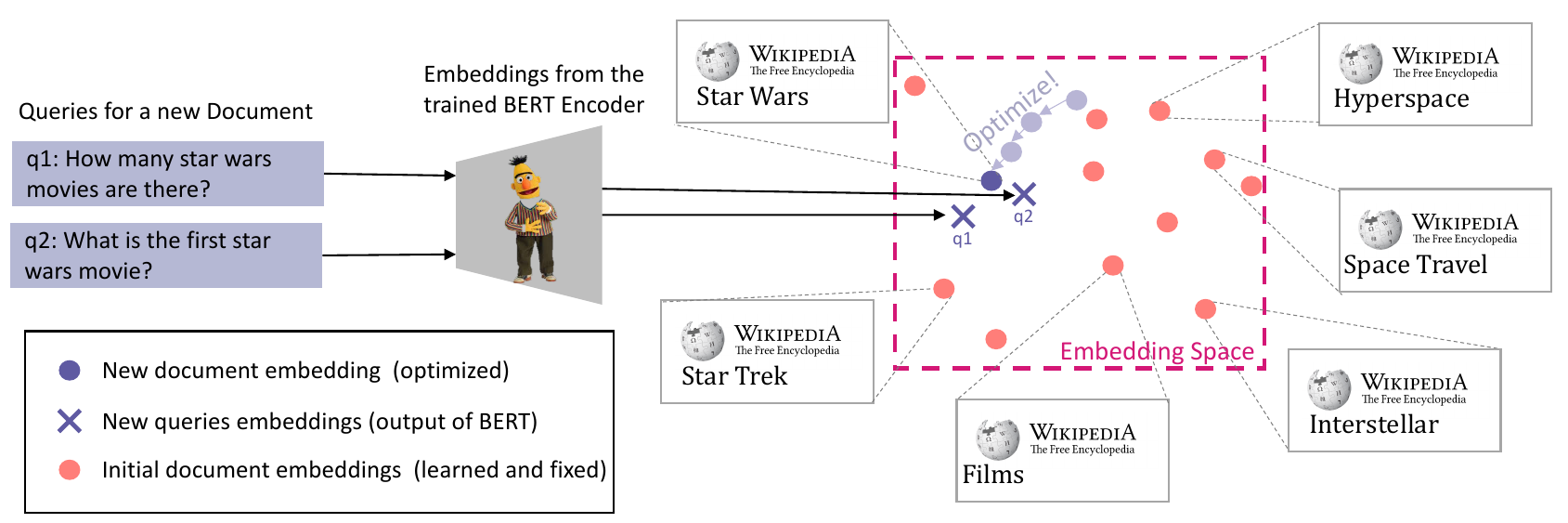}
    \caption{An illustration of the process of adding a new document (shown in purple) with its associated queries. The queries are embedded using the encoder trained on initial documents. A single document vector is optimized to be closer to the query embeddings (all other document vectors are fixed). 
}
    \label{fig:method1}
\end{figure*}

\section{IncDSI}
Before we introduce the constrained optimization problem used to obtain model $M^t$, we first introduce how the initial model $M^0$ is trained on the inital document corpus $D^0$. 

\subsection{Document Retrieval Model}
Our initial document retrieval architecture is a modified version of the DSI model~\cite{tay2022transformer}. As introduced in \autoref{sec:related_work}, DSI is a new end-to-end paradigm for document retrieval in which a single model is trained to directly produce the corresponding document id (docid) for a given query. DSI makes use of a T5 model backbone that is trained with either a language model head to autoregressively generate docids as strings or a classification layer to output atomic docids (atomic docids are arbitrary unique docids that are assigned to each document). We focus on the setup of a DSI model with a classification layer, as prior work~\cite{mehta2022dsi++} has shown that this approach is less prone to forgetting when compared to autoregressive methods. The atomic DSI network is trained with cross entropy loss to both index the documents and train the retrieval model. For indexing, the model is trained to map the first 32 tokens of a document to its corresponding docid, and for retrieval, it is trained to map user queries to corresponding docids. We make two main changes to the DSI as explained below.

For indexing, instead of using the first 32 tokens as document representation, we use an off-the-shelf query generation model like docTTTTTquery~\cite{nogueira2019doc2query} to generate queries for every document, and train the model to map the generated queries to corresponding docids. Prior work~\cite{wang2022neural,zhuang2022bridging} has demonstrated that using generated queries to index models yields better performance because it reduces the train-test gap between using extracted document text during training and user queries during test time. \citet{wang2022neural} obtain results by using both generated queries and first few document tokens. Our experiments suggest that using document tokens provide only minor benefits, so for simplicity we only use the generated queries to index the documents. We present these experiments in \autoref{app:doc_index}.

Since we are not using performing autoregressive decoding, we replace the T5 backbone (an encoder-decoder model) in DSI with a BERT backbone (an encoder model). Additionally, most dual encoder based methods are built using pre-trained BERT models~\cite{karpukhin2020dense,xiong2020approximate} and thus we can compare apples to apples by using a BERT. That said our method is invariant with respect to the choice of model; any other encoder (like encoder-only T5) can be used as well.

% DSI models tend to over-fit to the information in the training documents because they are designed to memorize the training documents. While this is desirable for indexing documents, it leads to poor generalization for new unseen documents. Dual encoder methods like DPR~\cite{karpukhin2020dense} that are pretrained with contrastive loss generalize better to un-seeen documents because contrastive learning produces robust representation spaces ~\cite{wortsman2022robust,radford2021learning,andreassen2021evolution}. To leverage the benefits of dual encoder methods, we replace the T5 backbone with a BERT query encoder that is pretrained with DPR\footnote{Generated queries are used to obtain document representations for pre-training DPR as well.}. In order to preserve the generalizable representation space learned by using DPR for pre-training, we freeze the parameters for the query encoder when finetuning our model. 

To summarize, our document retrieval model $M^0$, that is trained on the initial data $D^0$, consists of a BERT based query encoder and an additional classification layer. Akin to DSI, the model $M^0$ is trained with a cross entropy loss to perform classification and directly predict docids. 

\subsection{Incremental Addition}
The document retrieval model $M^0$ that is trained on documents $D^0$, has a classification layer $\mathbf{V} \in \mathbb{R}^{|D^0| \times h}$, where $|D^0|$ is the number of already indexed documents and $h$ is the output dimensionality of the query encoder. Each row in matrix $\mathbf{V}$ can be interpreted as a document vector (in $\mathbb{R}^h$) that corresponds to a particular document.
We can add a new document class to model $M^0$ by introducing an additional class vector corresponding to the new document to $\mathbf{V}$. To add the new document, we use the queries associated with that document and attempt to ensure that those queries are correctly mapped to new document. For reducing the mismatch between train and test time and to have a greater diversity of queries, we obtain additional queries with a query generation model as described in the previous section. In settings where natural queries are unavailable, just the generated queries can be used.

\paragraph{Optimization Problem.}
We formulate the addition of a new document, as a constrained optimization problem over the document representation space. We first describe how to add one new document to the model trained on the initial set and then describe how to use a similar procedure to add more documents sequentially.

Let's suppose that the current retrieval model has been trained on $n$ documents (so the number of rows in $\mathbf{V}$ is $n$). In order to add a new document $d_{n\!+\!1}$ with associated queries $\{\mathbf{q}_{0,{n\!+\!1}}, \cdots, \mathbf{q}_{k,{n\!+\!1}} \}$, we want to find some document representation $\mathbf{v}_{n\!+\!1} \in \mathbb{R}^{h}$ such that when $\mathbf{v}_{n\!+\!1}$ is appended to the existing classification layer $\mathbf{V}$, the resulting model \emph{both} correctly classifies queries corresponding to the new document and the documents that were 
already indexed when $M^0$ was trained. 

The first constraint we need to satisfy is to correctly classify any query from the new document. Because queries can be noisy, we average over the $k$ available queries to develop a representative query embedding $\bar{\mathbf{q}}_{n\!+\!1} = \frac{1}{k}\sum_{i=1}^k \mathbf{q}_{i,{n\!+\!1}}$ that should retrieve the new document 
($k$ is variable for every docuemnt). More formally, the constraint 
\begin{align}
\bar{\mathbf{q}}^T_{n\!+\!1}\mathbf{v}_{n\!+\!1} > \max_{1 \leq j \leq n} \bar{\mathbf{q}}^T_{n\!+\!1} \mathbf{v}_j
\label{eq:1}
\end{align}
should hold, where $\mathbf{v}_j$ is the $j$-th row of $\mathbf{V}$, $\bar{\mathbf{q}}^T_{n\!+\!1}\mathbf{v}_{n\!+\!1}$ is the score for the new document and $\bar{\mathbf{q}}^T_{n\!+\!1} \mathbf{v}_j$ is the score for the $j^\text{th}$ original document. The inequality in (\ref{eq:1}) ensures that the new document is scored higher than all the existing documents for the representative query embedding, and thus we ensure that the ``query" $\bar{\mathbf{q}}^T_{n\!+\!1}$ 
retrieves the new document.  

Although we want to retrieve the new document when appropriate, we do not want the addition of new documents to degrade retrieval performance for the original documents. So we need to minimize the probability that the queries corresponding to the original documents are also mapped to the new document. To achieve this, we use the set of queries used for indexing the original documents to introduce an additional set of constraints. For some original document $j$, we denote the cached set of training queries as $\{\mathbf{z}_{i, j}\}_{i=1}^k$; all training queries corresponding to the initial documents are cached after training the initial retrieval model $M^0$ for efficiency. We compute a representative query embedding by averaging over the cached queries $\bar{\mathbf{z}}_{j} = \frac{1}{k}\sum_{i=1}^k \mathbf{z}_{i, j}$. We can then construct a matrix $\mathbf{Z} \in  \mathbb{R}^{|D^0| \times h}$, that contains a representative query embedding for each original document.

To preserve the performance of our system for the original documents, we find a new class vector $\mathbf{v}_{n\!+\!1}$ that does not interfere with the retrieval of the existing documents. More formally, we enforce the following constraints
\begin{align}
\forall_j \mathbf{z}^T_j \mathbf{v}_{n\!+\!1} < \mathbf{z}^T_j \mathbf{v}_j\text{,}
\label{eq:2}
\end{align}
 where $\mathbf{z}_j$ is the $j$-th row of $\mathbf{Z}$, $\mathbf{z}^T_j \mathbf{v}_{n\!+\!1}$ is the score for the new document and $\mathbf{z}^T_j\mathbf{v}_j$ is the score for the $j^\text{th}$ original document. The inequalities in (\ref{eq:2}) ensure that the queries for each original document will not retrieve the new document.

Consequently, we find a $\mathbf{v}_{n\!+\!1}$ that correctly classifies old and new queries with the following optimization problem:
\begin{empheq}[box=\fbox]{align}
\min & \rVert \mathbf{v}_{n\!+\!1} \lVert_2^2 \nonumber\\
\text{s.t. } & \bar{\mathbf{q}}^T_{n\!+\!1}\mathbf{v}_{n\!+\!1} > \max_{1 \leq j \leq n} \bar{\mathbf{q}}^T_{n\!+\!1} \mathbf{v}_j, \nonumber \\
& \forall_j \mathbf{z}^T_j \mathbf{v}_{n\!+\!1} < \mathbf{z}^T_j \mathbf{v}_j.
\label{eq:3}
\end{empheq}

We rewrite the violation of the first constraint in a form amenable for optimization using the hinge loss
\begin{align}
\ell_{1}(\mathbf{v}_{n\!+\!1}) = \max(0, (\text{max}_j(\bar{\mathbf{q}}^T_{n\!+\!1}\mathbf{v}_j) - \bar{\mathbf{q}}^T_{n\!+\!1} \mathbf{v}_{n\!+\!1})) + \gamma_1 )^2,
\label{eq:4}
\end{align}
where $\gamma_1 > 0$ is some margin. We minimize the squared hinge loss because smooth variants of the hinge loss can be easier to minimize with first-order optimization methods \citep{Zhang2001, Rennie2005}. We found that this accelerated optimization while performing similarly to standard hinge loss. Minimizing \autoref{eq:4} satisfies the first constraint when the loss is low, finding some $\mathbf{v}_{n\!+\!1}$ that is retrieved by the new queries. We can also similarly rewrite the second constraint using the hinge loss as 
 \begin{align}
     \ell_2(\mathbf{v}_{n\!+\!1}) = \sum_j \max(0, \mathbf{z}^T_j \mathbf{v}_{n\!+\!1} - \mathbf{z}^T_j \mathbf{v}_j + \gamma_2 )^2,
     \label{eq:5}
 \end{align}
where $\gamma_2 > 0$ is some margin. Minimizing \autoref{eq:5} satisfies the second set of constraints when the loss is low and ensures that we find some $\mathbf{v}_{n\!+\!1}$ that does not interfere with the retrieval of the original documents.

Our final optimization objective is a convex combination of $\ell_1(\mathbf{v}_{n\!+\!1})$, which ensures that we retrieve the new document correctly, and $\ell_2(\mathbf{v}_{n\!+\!1})$, which ensures that we maintain performance for the old documents. Therefore our final optimization objective becomes 
\begin{equation*}
\begin{split}
    \mathcal{L}(\mathbf{v}_{n\!+\!1}) = \lambda_1 \ell_1(\mathbf{v}_{n\!+\!1}) +
    (1-\lambda_1) \ell_2(\mathbf{v}_{n\!+\!1}) + \lambda_2 \rVert \mathbf{v}_{n\!+\!1} \lVert_2^2,
\end{split}
\end{equation*}
where $\lambda_1\!\in\!(0,1)$ balances the objectives for accurately retrieving the new document and preserving the retrieval performance for the old documents, and $\lambda_2$ controls 
the weight for L2 regularization.
\begin{figure*}
    \centering
    \includegraphics[width=\linewidth]{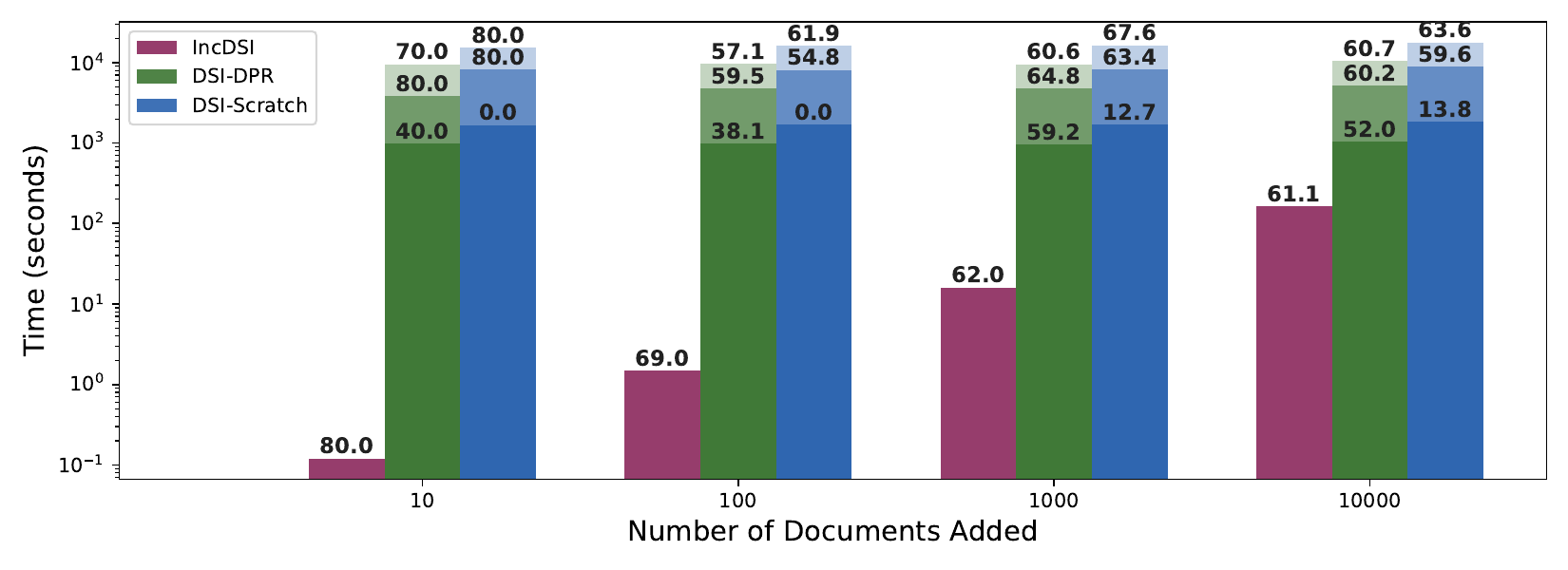}
    \caption{Time taken to add documents for different methods. Numbers on the bars are hit@1 for new documents. Lighter shades in stacked bars indicate later checkpoints (epochs 1,5,10). DPR, which only requires embedding queries and computing inner products, is not shown because it uses a model trained on just the original data and results in worse performance (when compared to the models here).}
    \label{fig:nq_time}
\end{figure*}

To solve the optimization problem, we utilize the L-BFGS optimizer \citep{fletcher2013practical}. For all of our experiments we set the initial learning rate to $1$ and utilize the strong Wolfe line search method \citep{NoceWrig06} to compute the step sizes during optimization. Both the L-BFGS optimizer and the strong Wolfe line search method are implmented natively within Pytorch. We optimize the weight vector for a maximum of $30$ iterations and terminate optimization early if the norm of the update is less than $10^{-3}$.

\paragraph{Algorithm.} So far, we have outlined how to add a single new document to a model trained some initial documents (see \autoref{fig:method1} for an overview). After finding $\mathbf{v}_{n\!+\!1}$ for the new document, we add it as a new row to matrix $\mathbf{V}$. We obtain an updated matrix of document representations $\mathbf{V}' = [\mathbf{V};\mathbf{v}_{n\!+\!1}] \in \mathbb{R}^{(|D|+1) \times h}$ and correspondingly an updated matrix of representative queries $\mathbf{Z}' = [\mathbf{Z};\bar{\mathbf{q}}] \in  \mathbb{R}^{|D| \times h}$, where $\bar{\mathbf{q}} = \frac{1}{m}\sum_{i=0}^m \mathbf{q}_i$ is the average query representation used to index the new document. We can now use the updated matrices $\mathbf{V}', \mathbf{Z}'$, treat the new document as a part of the already indexed initial document set and another new document. Therefore, we can repeatedly use the same optimization method described above to continue adding a stream of new documents. We outline this procedure for adding a new set of documents $D'$ in Algorithm~\ref{alg:IncDSI}.

\begin{algorithm}[tb]
   \caption{IncDSI}
   \label{alg:IncDSI}
\begin{algorithmic}
   \STATE {\bfseries Input:} query embeddings $\mathbf{Z}$, classification layer $\mathbf{V}$, new documnent set $D'$, new queries $\{q_{i, t}\}_{i=1}^k$ for every $t$-th new document ($k$ is variable for each docuemnt)
   \STATE {\bfseries Hyperparameters:} margins $\gamma_1$ and $\gamma_2$, loss weighting $\lambda_1$, l2 regularization weight $\lambda_2$
   \STATE n = number of initial rows in $\mathbf{V}$
   \FOR{document number $t$ {\bfseries in} $\{ 1, 2, \cdots, |D'|\}$}
   \STATE x=n+t
   \STATE Initialize $\mathbf{v}_{x}$ randomly
   \STATE $\bar{\mathbf{q}}_{x} = \frac{1}{n}\sum_i^n \mathbf{q}_{i,x}$
   \STATE $\text{optim} \gets \text{LBFGS}(\mathbf{v}_{x}, \text{lr}=1, \text{line\_search}=\text{True})$
   % \STATE $\text{optim} \gets LBFGS(w_d, \text{lr}=1, \text{line_search}=)$
   \REPEAT
   \STATE $\ell_{1}(\mathbf{v}_{x}) = \max(0, (\text{max}_j(\bar{\mathbf{q}}^T_{x}\mathbf{v}_j) - \bar{\mathbf{q}}^T_{x} \mathbf{v}_{x})) + \gamma_1 )^2$
   \STATE $\ell_2(\mathbf{v}_{x}) = \sum_j \max(0, \mathbf{z}^T_j \mathbf{v}_{x} - \mathbf{z}^T_j \mathbf{v}_j + \gamma_2 )^2$
   \STATE $\mathcal{L}(\mathbf{v}_{x}) = \lambda_1 \ell_1(\mathbf{v}_{x})+ (1-\lambda_1) \ell_2(\mathbf{v}_{x}) + \lambda_2 \rVert \mathbf{v}_{x} \lVert_2^2$
   \STATE step optim($\mathcal{L}(\mathbf{v}_{x})$) to minimize loss
   \UNTIL{30 iterations or $\rVert \Delta \mathbf{v}_{x} \lVert_2^2 < 10^{-3}$}
   \STATE $\mathbf{V} \gets [\mathbf{V};\mathbf{v}_{x}]$
   \STATE $\mathbf{Z} \gets [\mathbf{Z};\bar{\mathbf{q}}_{x}]$
   \ENDFOR
\end{algorithmic}
\end{algorithm}

\section{Experiments}
\label{sec:experiments}
\paragraph{Datasets.}
\label{sec:dataset}
We conduct our experiments on two publicly available datasets---Natural Questions 320K~\cite{kwiatkowski2019natural} and MS MARCO Document Ranking~\cite{nguyen2016ms}. We construct new benchmark datasets from Natural Questions and MS MARCO to facilitate research in building update-able document retrieval models.

The NQ320K dataset consists of query-document pairs, where the queries are natural language questions and the documents are Wikipedia articles that contain answers to the queries. MS MARCO is another popular question answering dataset that contains Bing questions and corresponding web page documents. The original dataset contains 3.2 million documents, but only a subset of these documents have associated queries. In each dataset, we assign a unique docid to each document.

% The two above-mentioned datasets have some documents that are unique to the test set. Since, we are directly training with a classification loss and want to ensure that no classes are `forgotten,` we add two generated queries for each document that is just in the training set to the test set. 

The documents in NQ320K and MS MARCO are each split into three sets---the initial document set $D^0$ that is available at the start, the new document set $D'$ that is available in a streaming fashion after a model is trained on the initial data and the tuning document set $D^*$ that is used to tune the parameters for \method{}. 
We randomly sample 90\% of the documents to form the initial set $D^0$, 9\% of the documents to form the new set $D'$ and 1\% of the documents to form the tuning set $D^*$. Each dataset also has natural human queries that are associated with the documents. We use the official NQ and MSMARCO train-validation splits to divide the queries into train/val/test splits as follows: the train split is divided into 80\% train/ 20\% validation data and the validation split is used as test data. For each document in the train set, 15 additional queries are generated using docTTTTTquery~\cite{nogueira2019doc2query}. Since query generation models sometimes produce the same generic query for multiple documents, we filter out queries that are linked to multiple documents. As a result, a few documents might have fewer than 15 generated queries. The final statistics of the two datasets is shown in \autoref{tab:statistics}.

\begin{figure*}
    \centering
        \centering
        \includegraphics[width=0.9\textwidth]{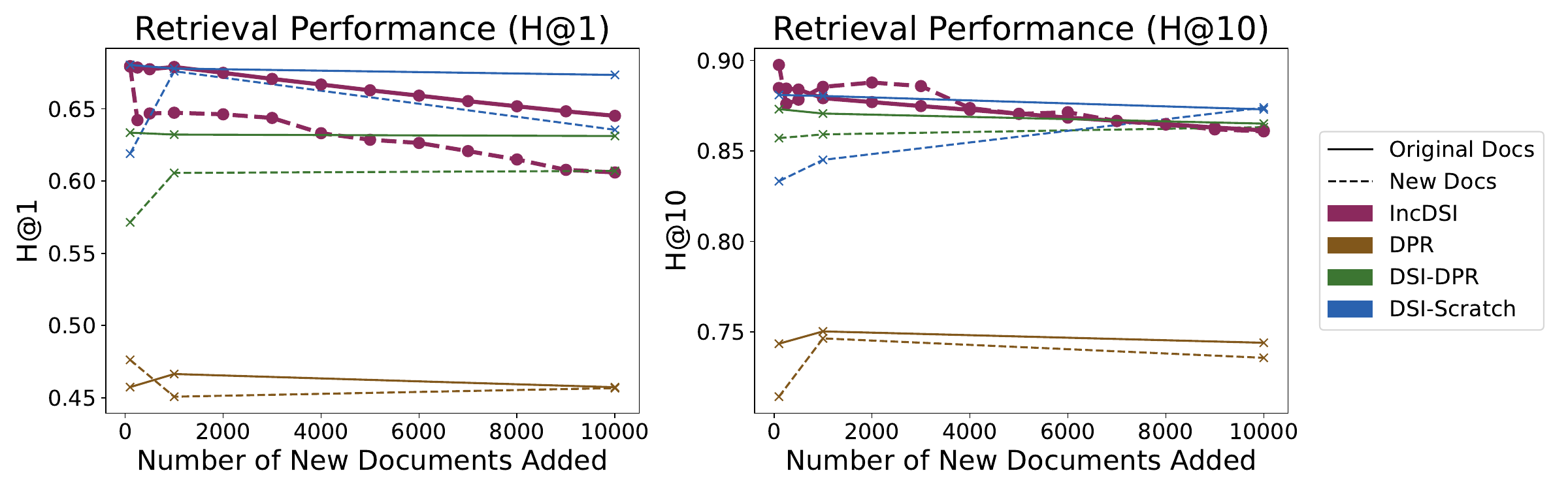} % second figure itself
        \caption{We present the retrieval performance for the original documents and new documents as increasing numbers of documents are indexed. The \method{} performance represents the average over 10 random document orderings. \label{fig:hits_curves}}
\end{figure*}

\paragraph{Baselines.} We compare our method with the following three baselines:
\label{sec:baselines}
\begin{itemize}
    \item DPR~\cite{karpukhin2020dense}: We train a standard dual-encoder DPR model on the initial dataset. The frozen encoder from the trained DPR model is used to obtain representations for documents and queries from the original and new dataset. Nearest neighbor search is then used to classify the queries. 
    \item Continual training with frozen DPR (DSI-DPR): A model consisting of a frozen DPR encoder and a trainable classification layer is continually fine-tuned with cross-entropy loss on natural and generated queries from both the old and new documents. 
    % The same DPR checkpoint is used for the baselines and for \method. 
    \item Continual training (DSI-Scratch): A DSI model is first trained to map generated and natural queries from the initial documents to their corresponding docids (to make a fair comparison, we use BERT as the backbone for the DSI model). The model is then continually fine-tuned with queries from the old and new documents. This method is similar to DSI++~\cite{mehta2022dsi++}. During continual training, we utilize the same hyperparameters as the model trained on old documents. 

\end{itemize}

\paragraph{Experimental Setting.} 
We use the BERT model~\cite{devlin2018bert} and initialize it with publicly available bert-base-uncased weights for all our experiments. The classification layer is randomly initialized. For the DPR baseline, we use the offcial implementation~\cite{karpukhin2020dense}. For the continual training baselines, the document retrieval model is trained for 20 epochs on the initial set of documents and for an additional 10 epochs on both the initial and new documents. A learning rate of 1e-5 and 5e-5 and a batch size of 128 and 1024 are used for NQ320K and MSMARCO respectively. The results are reported for the epoch with the best validation accuracy.
 For all our experiments, we use one A6000 GPU.

\paragraph{Metrics.} In line with previous work~\cite{tay2022transformer,wang2022neural}, we measure Hits@k, where $k=\{1,5,10\}$, and Mean Reciprocal Rank@10 (MRR@10) to evaluate our method and the baselines.  Hits@k (also denoted as H@k) measures how often the desired document is one of the top-k retrieved documents and MRR@k calculates the reciprocal of the rank at which the correct document is retrieved (the rank is set to infinity if the desired document is not in the top $k$). We measure these metrics on both queries belonging to the initial documents $D^0$ and the newly added documents $D'$. We also measure the amount of time required to add the new documents to an already trained retrieval model.

\paragraph{Hyperparameter tuning.}
To tune the four hyperparameters for \method{} (that is objective trade-off weight $\lambda_1$, L2 regularization weight $\lambda_2$ and margins $\gamma_1, \gamma_2$), we utilize the Ax library \citep{Bakshy2018AEAD} to perform bayesian optimization with our tuning set $D^*$. We run hyperparameter optimization for 50 trials with the default Ax library settings and use the best hyperparameters to add the heldout documents in the new document set $D'$. We optimize the hyperparamters over ${\lambda_1 \in \text{Uniform}(.05,.95)}$, ${\gamma_1, \gamma_2 \in \text{Uniform}(0,10)}$, and ${\lambda_2 \in \text{LogUniform}(1e-8,1e-3)}$. For bayesian optimization, we set the target objective to the F-beta score and compute the weighted harmonic mean of the validation MRR@10 for the original documents and the tuning documents. Formally, given the validation MRR@10 for the original documents, $y_{\text{orig}}$, and the validation MRR@10 for the tuning documents, $y_{\text{tune}}$, the target metric is 
\begin{equation*}
    y_{\text{target}} = (1+\beta^2)\frac{y_{\text{tune}} \cdot y_{\text{orig}}}{(\beta^2 \cdot y_{\text{tune}}) + y_{\text{orig}}},
\end{equation*}
where setting $\beta > 1$ emphasizes the retrieval performance for the original documents. Because the the original document set is generally much larger than the set of new documents, we set $\beta=5$ to emphasize the preservation of the retrieval performance for the existing documents (this choice is ablated in the next section). The target objective can be modified to include time or hits@k information depending on the specific use case.

\section{Results and Discussions.}
% \paragraph{Generated queries only.}
% The results on our method and other baselines are shown in \ref{}. 
\paragraph{Performance.}
\label{sec:results}
We add $k$ documents, where $k \in \{10,100,1000,10000\}$, to a model trained with the initial documents $D^0$ and evaluate the retrieval accuracy and time required to add documents with \method{} and the baselines introduced previously. We empirically observe that every new document can be added by satisfying all the constraints in \autoref{eq:3} for the datasets we use. However, there might exist cases when the optimization problem fails to find a feasible solution. In such a case, the optimization problem can be re-started after altering it by tweaking the initialization/hyperparameters or by removing some queries and re-computing the representative query $\bar{\mathbf{q}}$ in \method{}. 

\autoref{fig:nq_time} and \autoref{fig:hits_curves} show time and accuracy plots for a adding different number of documents with the NQ320K dataset. The raw numbers are presented in \autoref{app:nq_results} and \autoref{app:marco_results}. The trends for the MS MARCO dataset are similar and due to space constraints the results on the MS MARCO dataset are presented in \autoref{app:marco_results}. 

With \method{}, we can add previously unseen documents to the index in less than 50 milliseconds. This means that our approach can efficiently index a stream of documents as they become available. We observe that retraining the DSI models takes orders of magnitudes longer to achieve comparable performance on the new documents. For example, \method{} indexes 1000 documents in roughly 16 seconds and achieves a H@1 of 62.0 for those documents. The baseline DSI-Scratch, on the other hand, needs over $513\times$ longer (2hr17m) to achieve the H@1 of 63.4.  Moreover, on the MS MARCO dataset, \method{} outperforms the baselines despite requiring orders of magnitude lesser time; \autoref{tab:8} shows that H@1 for the new documents is 61.0 with \method{} and at most 51.8 for the baselines. The learned baselines need to be trained for much longer than 10 epochs (which already takes about 6 hours) to achieve better performance on the new documents because the initial MS MARCO document set is much bigger. As a result, using such methods in a streaming setting is impractical. Our constrained optimization formulation, however, is able to find an effective new document representation in a fraction of a second.  

Compared to the dual-encoder DPR baseline that can also encode new documents in a streaming setting in milliseconds, we observe that our method is similarly fast while consistently achieving greater retrieval performance (see \autoref{tab:7} and \autoref{tab:8}). For all settings with a reasonable sample size (i.e. $\geq100$ document additions), \method{} on NQ320K achieves a H@1 greater than 61.0 on the new documents while DPR never exceeds 48.0. This is due to leveraging the strengths of DSI and decoupling the document representations from a parametric model like a BERT encoder. By directly optimizing over the representation space, our model has much greater capacity to incorporate information from new documents.

We present the retrieval performance of our system in a streaming setting where documents are added incrementally to the index in \autoref{fig:hits_curves}. Our approach is capable of indexing thousands of documents effectively with limited interference with the original documents. Notably, the H@10 for the original documents is nearly constant during indexing, although the Hits@1 for the original documents does degrade slowly over time.

These results show that \method{} is very effective for adding documents in close to real time and yields close to the same performance as retraining. While \method{} is not a replacement for the standard paradigm of retraining, particularly in settings where many documents must be added to the index, it offers a solution for indexing documents in real-time and can potentially reduce the frequency at which resource-intensive retraining is required.

\paragraph{Using only generated queries.}
There are scenarios where no natural queries are available for new documents. For instance, when a new paper is uploaded to arXiv, human queries corresponding to that paper might not be available. In this scenario, we can add documents by using only generated queries. In table \autoref{tab:gen_only}, we report the performance from using only generated queries to add 1000 new documents and we see that \method{} achieves comparable performance to the baselines. When these numbers are compared to those obtained from using both natural and generated queries, we observe that significant gains are achieved by using natural queries. This is likely because natural queries are more diverse in structure and content than generated queries from the docTTTTTquery generation model. Using a better query generation model will help improve performance, especially when only using generated queries.

\begin{table}
\centering
\caption{Impact of only using generated queries}\label{tab:gen_only}
\resizebox{\columnwidth}{!}{%
\begin{tabular}{lcccccc}
\toprule
  & \multicolumn{5}{c}{NQ320K (Original/New)}  \\
  \cmidrule{2-6}
    & H@1 & H@5 & H@10 & MRR@10 & Time (min) \\
    \midrule
  \method{} & $67.7/53.5$ & $84.6/77.5$ & $87.9/81.7$ & $75.1/63.7$ & 13.85s \\
  DSI-DPR & 63.4/53.5 & 83.3/74.6 & 87.3/78.9 & 72.0/61.8 & 45m12s \\
  DSI-Scratch & 68.0/53.5 & 84.4/76.1 & 87.7/78.9 & 75.2/62.7 & 215m36s \\
\bottomrule
\end{tabular}
}
\end{table}

\paragraph{Ablations.} We ablate a number of the different design choices made in developing our framework. For our ablation studies, we report results for adding 1000 new documents from the NQ320k dataset.

\begin{table}
\centering
\caption{Impact of different values of $\beta$.}\label{tab:beta_ablation}
\resizebox{\columnwidth}{!}{%
\begin{tabular}{lccccc}
\toprule
  & \multicolumn{4}{c}{NQ320K (Original/New)}  \\
  \cmidrule{2-5}
    & H@1 & H@5 & H@10 & MRR@10 \\
    \midrule
  $\beta=1$ & $65.4/76.1$ & $83.6/85.9$ & $87.3/87.3$ & $73.4/80.3$ \\
  $\beta=3$ & $67.3/67.6$ & $84.4/81.7$ & $87.7/84.5$ & $74.7/73.7$ \\
  $\beta=5$  & $67.8/63.4$ & $84.8/73.2$ & $88.0/76.1$ & $75.2/68.1$ \\
  $\beta=10$ & $68.2/52.1$ & $84.9/66.2$ & $88.1/70.4$ & $75.5/58.5$ \\
\bottomrule
\end{tabular}%
}
\end{table}

\paragraph{The Bayesian optimization target.} We ablate the impact of the weighting term, $\beta$, used during hyperparameter tuning. Increasing $\beta$ places more emphasis on maintaining retrieval performance for the original documents. We report results over a sweep of different $\beta$ values in \autoref{tab:beta_ablation}. As expected, increasing $\beta$ monotonically improves the performance on the original documents at the expense of performance on the new documents. We selected $\beta=5$ as it strikes a reasonable balance, but different values may be adviseable depending on the application.

\begin{table}
\centering
\caption{Impact of using a different number of generated queries.}\label{tab:gen_query_ablation}
\resizebox{\columnwidth}{!}{%
\begin{tabular}{lcccc}
\toprule
  & \multicolumn{4}{c}{NQ320K (Original/New)}  \\
  \cmidrule{2-5}
   Num Queries & H@1 & H@5 & H@10 & MRR@10 \\
    \midrule
  $5$ & $68.1/56.3$ & $84.7/71.8$ & $87.9/80.3$ & $75.4/64.3$ \\
  $10$ & $68.0/60.6$ & $84.8/77.5$ & $87.9/78.9$ & $75.3/66.9$ \\
  $15$ & $67.8/62.0$ & $84.6/76.1$ & $87.9/81.6$ & $75.1/68.9$ \\
\bottomrule
\end{tabular}%
}
\end{table}

\paragraph{Number of generated queries.} We ablate the number of generated queries used in \autoref{tab:gen_query_ablation}. The results show that \method{} is not sensitive to the number of generated queries for NQ320k. There might be other datasets that benefit from a greater number of generated queries due to greater diversity. Using many generated queries leads to obtaining a more robust representation of the document class that leads to better generalization. 

\begin{table}[h]
\centering
\caption{Impact of loss function.}\label{tab:loss_ablation}
\resizebox{\columnwidth}{!}{%
\begin{tabular}{lccccc}
\toprule
  & \multicolumn{4}{c}{NQ320K}  \\
  \cmidrule{2-6}
   Loss Function & H@1 & H@5 & H@10 & MRR@10 & Time\\
    \midrule
  Hinge & $68.1/59.2$ & $84.8/74.6$ & $88.0/80.3$ & $75.4/66.4$ & 53.7s \\
  Squared Hinge & $67.8/62.0$ & $84.6/76.1$ & $87.9/81.6$ & $75.1/68.9$ & 16.1s \\
\bottomrule
\end{tabular}%
}
\end{table}

\paragraph{Loss function.} We report the effect of minimizing the standard hinge loss instead of the squared hinge loss in \autoref{tab:loss_ablation}. We observe that they achieve similar performance. However, using the squared hinge loss is almost $3.3 \times$ faster and we therefore use the squared hinge loss as the loss function of \method{}.

\section{Limitations and Future work}
Despite enabling a real time document retrieval system with good retrieval accuracy, our method has some limitations. As we add an increasing number of new documents, the performance on the original set of documents degrades slightly, and we eventually need to retrain the model (as is standard practice). It is possible that alternative formulations of the optimization objective would be more effective at preserving performance for longer.

To embed new queries, we use a frozen query encoder that is trained on the set of initial documents and thus rely on strong representations from the query encoder to generalize effectively. In the future, we would like to explore pretraining tasks or other methods to improve the generalizability of the query encoder.
We can also further improve the performance of \method{} by training a query generation model on in-domain data, instead of using an off-the-shelf model. 

In this work we only consider the setting of adding new documents. However, our proposed method can also be used to edit information in existing documents. If we want to edit the information in document, we can formulate new constraints that encode the information that needs to be associated with the document and optimize its corresponding document vector using \method{}. We leave the exploration of editing docuemnts to future work.
% add something about string dsi

\section{Conclusion}
We present, \method{}, a novel document retrieval system that can index new documents as soon as they are available in roughly 50 milliseconds. We accomplish this by formulating the problem of indexing a new document as a constrained optimization problem over the document representation space. By holding the rest of our system fixed and optimizing only the document representation, we can rapidly introduce new documents to our system. \method{} is orders of magnitudes faster when compared to retraining the document retrieval model and yet produces comparable performance.

\section*{Acknowledgements}
This research is supported by a gift from the Simons Foundation, and grants from DARPA AIE program, Geometries of Learning
(HR00112290078),
the National Science Foundation NSF (IIS-2107161, and IIS-1724282, HDR-2118310, OAC-2118310),
and the Cornell Center for Materials Research with funding from the NSF MRSEC program (DMR1719875). We thank Oliver Richardson, Katie Luo and all the reviewers for their feedback.


\bibliography{paper}
\bibliographystyle{icml2023}

%%%%%%%%%%%%%%%%%%%%%%%%%%%%%%%%%%%%%%%%%%%%%%%%%%%%%%%%%%%%%%%%%%%%%%%%%%%%%%%
%%%%%%%%%%%%%%%%%%%%%%%%%%%%%%%%%%%%%%%%%%%%%%%%%%%%%%%%%%%%%%%%%%%%%%%%%%%%%%%
% APPENDIX
%%%%%%%%%%%%%%%%%%%%%%%%%%%%%%%%%%%%%%%%%%%%%%%%%%%%%%%%%%%%%%%%%%%%%%%%%%%%%%%
%%%%%%%%%%%%%%%%%%%%%%%%%%%%%%%%%%%%%%%%%%%%%%%%%%%%%%%%%%%%%%%%%%%%%%%%%%%%%%%
\newpage
\appendix
\onecolumn
\section{Datasets.}

Below are the statistics for our splits of the NQ dataset and the MS Marco dataset. 

\begin{table}[h]
\centering
\caption{Statistics of NQ320k and MSMARCO.}\label{tab:statistics}
\begin{tabular}{lccccc}
\toprule
  & \multicolumn{5}{c}{NQ320K}  \\
  \cmidrule{2-6}
    & Document & Train & Val & Test & Generated queries \\
    \midrule
  $D$ & 98743 & 221194 & 55295 & 6998 & 1480538 \\
  $D'$ & 9874 & 22178 & 5545 & 738 & 148077 \\
  $D*$ & 1098 & 2525 & 632 & 94 & 16470 \\
\bottomrule
\end{tabular}

\begin{tabular}{lccccc}
\toprule
  & \multicolumn{5}{c}{MSMARCO}  \\
  \cmidrule{2-6}
    & Document & Train & Val & Test & Generated queries \\
    \midrule
  $D$ & 289424 & 262008 & 65502 & 4678 & 4312150  \\
  $D'$ & 28943 & 26197 & 6550 & 455 & 431222 \\
  $D*$ & 3216 & 2968 & 742 & 38 & 47921 \\
\bottomrule
\end{tabular}
\end{table}

Note that we use document titles to de-duplicate documents in the NQ320K. Some past work~\cite{tay2022transformer} has used URLs to de-duplicate documents, but because two different versions of a document (with different URLs) only vary in minor edits, this will lead to essentially the same document being mapped to two different document ids. 

As mentioned \autoref{sec:experiments}, for each document we produce generated queries using a docTTTTTquery model~\cite{nogueira2019doc2query}. To ensure the diversity, we sample queries by using different sections (of length 512 tokens) of the documents as input to docTTTTTquery. 

After adding $t$ new documents, the accuracy metrics for the new documents are computed by only using queries in the val/test set that correspond to one of the newly added documents.  

\section{Training with and without Document Text}
\label{app:doc_index}
\begin{table}[h]
\centering
\caption{Impact of training with document text}\label{tab:doc_text}
\begin{tabular}{lccccc}
\toprule
  & \multicolumn{4}{c}{NQ320K}  \\
  \cmidrule{2-5}
    & H@1 & H@5 & H@10 & MRR@10 \\
    \midrule
  DSI-DPR (w/ doc text) & $65.4$ & $83.8$ & $87.4$ & $72.3$  \\
  DSI-DPR (w/o doc text) & 64.1 & 83.2 & 87.4 & 71.8  \\
  DSI-Scratch (w/ doc text) & $68.4$ & $85.5$ & $88.5$ & $75.9$  \\
  DSI-Scratch (w/o doc text) & 68.0 & 85.3 & 88.4 & 75.6  \\
\bottomrule
\end{tabular}
\end{table}
We conducted experiments to investigate the impact of just using just the queries (generated and natural) versus using the queries plus the first $32$ tokens of the document to train the initial retrieval model. It can be seen from \autoref{tab:doc_text} that using the document text has a limited impact for DSI-scratch. It does moderately increase the Hits@1 DSI-DPR but has a limited impact on the other metrics. Therefore, for simplicity, we opted to just use just the queries to train the retrieval models.

\section{Results on NQ320K}
\label{app:nq_results}
Retrieval accuracy and time spent to add documents with \method{} and other baselines on are shown in \ref{tab:NQresults}. Here, DSI\-Scratch$_n$ refers to the model with a Bert encoder and a classification layer trained from scratch for $n$ epochs. DSI-DPR$_n$ denotes a model with a frozen pretrained DPR Bert encoder and a classification layer that is trained for $n$ epochs. We can see that the hits@1 (H@1) and hits@10 (H@10) of IncDSI are comparable to DSI-Scratch and DSI-DPR, while the time required to add new documents is much shorter for \method{}. The performaance of DPR is much worse because a frozen model trained on the initial data is used to encode the new documents and the new queries. The time is not shown for DPR because no additional training is required when adding the new documents with DPR.
\begin{table}
\centering
\caption{Hits@k of queries corresponding to original and new documents, and time (min) spent for adding different number of new documents}\label{tab:NQresults}
\resizebox{\columnwidth}{!}{
\begin{tabular}{c|ccccccccc}
\toprule
   \multicolumn{10}{c}{NQ320k} \\
  \multicolumn{10}{c}{Accuracy (Original/New)} \\
\midrule
Documents & Metric & IncDSI & DSI-Scratch$_1$ & DSI-Scratch$_5$ & DSI-Scratch$_{10}$ & DSI-DPR$_1$ & DSI-DPR$_5$ & DSI-DPR$_{10}$ & DPR \\
\hline
10 & H@1 & 68.0/80.0 & 68.2/0.0 & 68.1/80.0 & 67.9/80.0 & 63.8/40.0 & 63.1/80.0 & 63.4/70.0 & 46.7/60.0 \\
& H@10 & 88.5/80.0 & 88.5/40.0 & 88.3/80.0 & 88.1/80.0 & 87.5/80.0 & 87.4/80.0 & 87.3/80.0 & 75.1/90.0 \\
& time (min) & 0.002 & 27.9 & 136.4 & 256.0 & 16.4 & 63.7 & 158.7 & - \\
\hline
100 & H@1 & 67.9/69.0 & 68.2/0 & 68.2/54.8 & 68/61.9 & 63.8/38.1 & 63.4/59.5 & 63.4/57.1 & 45.7/47.6 \\
& H@10 & 88.5/85.7 & 88.4/19.1 & 88.3/76.2 & 88.1/83.3 & 87.4/69.1 & 87.4/81 & 87.3/85.7 & 74.4/71.4 \\
& time (min) & 0.03 & 28.0 & 135.5 & 268.1 & 16.5 & 80.0 & 160.0 & - \\
\hline
1000 & H@1 & 67.8/62.0 & 68.2/12.7 & 67.7/63.4 & 67.8/67.6 & 64.3/59.2 & 63.4/64.8 & 63.2/60.6 & 46.6/45.1 \\
& H@10 & 87.9/81.7 & 88.2/35.2 & 88.1/81.7 & 88/84.5 & 87.5/80.3 & 87.2/81.7 & 87.1/85.9 & 75/74.6 \\
& time (min) & 0.27 & 28.2 & 136.6 & 274.1 & 16.1 & 79.2 & 159.2 & - \\
\hline
10000 & H@1 & 64.4/61.1 & 68/13.8 & 67.4/59.6 & 67.3/63.6 & 63.7/52 & 63.1/60.2 & 63.1/60.7 & 45.7/45.7 \\
& H@10 & 86.3/86.3 & 87.9/35 & 87.6/85.2 & 87.3/87.4 & 86.7/84.4 & 86.3/87.4 & 86.5/86.3 & 74.4/73.6 \\
& time (min) & 2.73 & 30.6 & 148.5 & 297.3 & 17.5 & 86.9 & 175.2 & - \\
\bottomrule
\end{tabular}
}
\label{tab:7}
\end{table}

\section{Results on MS MARCO}
\label{app:marco_results}
The results on MS MARCO dataset is shown in \autoref{tab:MSMarcoresults}. Again, DSI\-Scratch$_n$ refers to the model with a Bert encoder and a classification layer trained from scratch for $n$ epochs. DSI-DPR$_n$ denotes a model with a frozen pretrained DPR Bert encoder and a classification layer that is trained for $n$ epochs. MS MARCO is harder than NQ320K because there are three times as many documents, so the retrieval accuracy is lower when compared to the NQ320k dataset for \method{} and the baselines. For this dataset, \method{} \emph{outperforms} the trained baselines (especially for performance of new documents). This shows that \method{} is especially good for making small updates to model trained on large datasets. The time is not shown for DPR because no additional training is required when adding the new documents with DPR.
\begin{table}[h!]
\centering
\caption{Hits@k of queries corresponding to original and new documents, and time (min) spent for adding different number of new documents}\label{tab:MSMarcoresults}
\resizebox{\columnwidth}{!}{
\begin{tabular}{c|ccccccccc}
\toprule
   \multicolumn{10}{c}{MSMARCO} \\
  \multicolumn{10}{c}{Accuracy (Original /New)} \\
\midrule
Documents & Metric & IncDSI & DSI-Scratch$_1$ & DSI-Scratch$_5$ & DSI-Scratch$_{10}$ & DSI-DPR$_1$ & DSI-DPR$_5$ & DSI-DPR$_{10}$ & DPR \\
\hline
10 & H@1 & 47.6/50.0 & 47.81/0 & 47.01/0 & 46.17/0 & 47.1/0 & 47.2/0 & 46.9/0 & 36.5/0 \\
& H@10 & 81.4/50.0 & 81.25/0 & 79.37/0.5 & 76.9/0.5 & 80.3/0 & 80.3/0.5 & 80.2/0.5 & 67.2/0.5 \\
& time (min) & 0.007 & 61.1 & 312.4 & 625.1 & 44.3 & 218.1 & 434.4 & - \\
\hline
100 & H@1 & 47.6/66.7 & 47.78/0 & 47.13/41.67 & 46.19/33.3 & 46.9/0 & 47.1/26.2 & 47/41.7 & 36.5/20.8 \\
& H@10 & 81.4/83.3 & 81.38/20.83 & 79.42/83.33 & 76.68/83.3 & 80.4/0 & 80.3/44.8 & 80.2/57.3 & 67.2/66.7 \\
& time (min) & 0.07 & 69.7 & 349.2 & 698.0 & 44.1 & 217.1 & 433.0 & - \\
\hline
1000 & H@1 & 49.7/65.5 & 49.8/6.9 & 49.1/55.2 & 47.9/55.2 & 48.8/0 & 48.8/37.9 & 48.9/44.8 & 39.5/31 \\
& H@10 & 81.4/82.8 & 81.5/27.6 & 79.8/75.9 & 76.6/79.3 & 80.7/0 & 80.9/72.4 & 80.5/75.9 & 69.1/58.6 \\
& time (min) & 0.68 & 70.0 & 352.8 & 698.3 & 42.7 & 207.6 & 414.2 & - \\
\hline
10000 & H@1 & 47.5/61.0 & 49.9/9.8 & 48.2/51.8 & 47.6/51.2 & 49.2/1.2 & 48.9/48.2 & 48.1/50.6 & 39.2/42.1 \\
& H@10 & 80.2/86.0 & 81.6/26.8 & 79.8/80.1 & 76.8/81.7 & 80.7/21.3 & 80.6/79.9 & 80.3/81.7 & 68.8/62.8 \\
& time (min) & 6.9 & 137.0 & 675.4 & 1372.7 & 46.3 & 224.8 & 448.9 & - \\
\bottomrule
\end{tabular}
}
\label{tab:8}
\end{table}

%%%%%%%%%%%%%%%%%%%%%%%%%%%%%%%%%%%%%%%%%%%%%%%%%%%%%%%%%%%%%%%%%%%%%%%%%%%%%%%
%%%%%%%%%%%%%%%%%%%%%%%%%%%%%%%%%%%%%%%%%%%%%%%%%%%%%%%%%%%%%%%%%%%%%%%%%%%%%%%

\end{document}